\documentstyle[12pt,fleqn,cite,epsfig]{article}
\newcommand{\sect}[1]{\setcounter{equation}{0}\section{#1}}
\newcommand{\subsect}[1]{\subsection{#1}}

\begin{document}
\begin{titlepage}
\begin{center}
\hfill IP/BBSR/2002-20\\
\hfill hep-th/0208042\\
\vskip .2in

{\Large \bf $p-p^{\prime}$ Branes in  PP-wave Background}
\vskip .5in

{\bf Anindya Biswas \footnote{e-mail: anindyab@iopb.res.in},
Alok Kumar\footnote{e-mail: kumar@iopb.res.in},  
and Kamal L. Panigrahi\footnote{e-mail: kamal@iopb.res.in}\\
\vskip .1in
{\em Institute of Physics,\\
Bhubaneswar 751 005, INDIA}}
\end{center}
\vskip 1.2in
\begin{center} {\bf ABSTRACT}
\end{center}
\begin{quotation}\noindent
\baselineskip 10pt
We  present several supergravity solutions corresponding to 
both $Dp$, as well as $Dp-Dp^{\prime}$ systems, in NS-NS and R-R
PP-wave background originating from $AdS_3 \times S^3\times R^4$. 
The $Dp$ brane solutions, $p=1,..,5$ are fully localized, whereas 
$Dp-Dp^{\prime}$ are localized along common transverse directions. 
We also discuss the supersymmetry properties of
these solutions and the worldsheet construction for the 
$p-p^{\prime}$ system.
\end{quotation}
\vskip 2in
August 2002\\
\end{titlepage}
\vfill
\eject
\sect {\bf Introduction}
PP-waves\cite{penrose,tsyt1,guven,metsaev,blau,tsyt2,malda,mukhi,GO,tseytlin,
sonn,mohsen,chu,lee,Bala,Taka} 
are known to be an interesting  class of supersymmetric 
solutions of type IIB supergravity, with wide applications 
to gauge theories. For these applications, 
such solutions are considered in the ``Penrose'' limit  of strings in 
$AdS_p\times S^p$ background\cite{blau}. In several cases, 
PP-waves are also  known to be maximally supersymmetric 
solutions of supergravities in 
various dimensions\cite{meessen,pope,hull}, and are known to  
give rise to solvable string theories from worldsheet 
point of view as well. The particular case of  
$AdS_5\times S^5$ is of special interest, with applications to 
$N=4$, $D=4$ gauge theories in the limit of large conformal 
dimensions and R-charges\cite{malda}. Interestingly, the $D$-branes 
of string theories also have an appropriate representation in 
such gauge theories, in terms of operators corresponding to 
`giant gravitons' and `defects'\cite{lee,Bala}. 

In this paper, we continue the search for explicit $D$-brane 
supergravity solutions in string theories in PP-wave background. 
Our study will be mainly concentrated to the NS-NS and R-R PP-waves 
arising out of $AdS_3\times S^3$ geometry \cite{tseytlin}. 
We, however, also present branes in other PP-waves such as the ones in 
`little' string theories etc.. 
Explicit supergravity solution of $D$-branes,
along with open string spectrum, has been  
studied in \cite{dabh,kumar,sken,alishah,bain,singh,pal}.
Our new result includes a set of solutions corresponding to 
the brane systems of the type: $Dp-Dp^{\prime}$ in both
IIA and IIB theories, as well as explicit worldsheet construction 
in  these cases. We would like to mention that 
$Dp$-branes from worldsheet point of view have already been 
obtained in NS-NS PP-wave background earlier \cite{michi}. Our result 
gives realization of such $D$-branes from the supergravity point of
view. We also find it interesting to note that, unlike the case of 
$AdS_5\times S^5$ PP-wave, in our case we are able to obtain several 
`localized' $D$-brane solutions. $Dp$-branes in R-R PP-wave background
are obtained by applying a set of $S$ and $T$-duality transformations
on the brane solutions in NS-NS background. Such background PP-wave 
configurations were already discussed in the context of $D7$-branes in 
\cite{kumar}. The worldsheet construction of these branes  
then follows from the results in \cite{kumar}, and will also be 
discussed below. We, however, point out that there is a crucial 
difference between the solutions presented in this paper and 
those in \cite{kumar}. In the present case, only light-cone 
directions of the PP-wave are along the branes, 
the remaining four are always transverse to them. On the other hand, 
for the solutions in \cite{kumar} the directions transverse to the
brane are flat.

The rest of the paper is organised as follows. In section-2, 
new (supersymmetric) $Dp$ as well as $Dp-Dp^{\prime}$ branes 
are presented in both NS-NS and R-R PP-wave backgrounds.
Supersymmetry properties of these branes have been examined in 
detail in section-3 and it is shown that they preserve
some amount of unbroken supersymmetry. 
Open string constructions of branes is discussed in section-4. 
Section-5 is devoted to the branes in `little string theory' 
background. We conclude in section-6 with some 
general remarks.  

\sect{\bf Supergravity Solutions}

\subsect{\bf $D$-Branes in NS-NS PP-wave background}

We now start by writing down the $D$-string solutions in the PP-wave 
background originating from the Penrose limit of NS-NS  
$AdS_3\times S^3\times R^4$ \cite{tseytlin}. 
The supergravity solution of a system of $N$ $D$-strings in such a 
background is given by:
\begin{eqnarray}
ds^2&=&f^{-{1\over2}}_1(2 dx^+dx^- - \mu^2{\sum_{i=1}^{4}}x_i^2(dx^+)^2)
+ f^{1\over2}_1{\sum_{a=1}^{8}}(dx^a)^2  \cr
& \cr
e^{2\phi}&=& f_1,~~~~H_{+12} = H_{+34}= 2\mu, \cr
& \cr
F_{+-a}&=&\partial_a f^{-1}_1,~~~~f_1 = 1 + {N g_s l^6_s\over r^6},
\label{d1-ads3}
\end{eqnarray}
with $f_1$ satisfying the Green function equation in the
8-dimensional transverse space.
We have explicitly verified that the solution presented in
(\ref{d1-ads3}) satisfy the type IIB field equations
(see e.g.\cite{duff,oz}). One notices that in this case,
constant NS-NS 3-forms along the PP-wave direction are
required precisely to cancel the $\mu$-dependent part of $R_{++}$
equation of motion.

Starting from the $D$-string solution in eqn. (\ref{d1-ads3}), one 
can write down all the $Dp$-brane solutions ($p=1,..,5$)
in NS-NS PP-wave background by applying succesive T-dualities 
along $x^5, ..,x^8$. As is known, this procedure also involves
smearing of the brane along these directions. For example, 
a $D3$-brane solution has a form: 
\begin{eqnarray}
ds^2&=&f^{-{1\over2}}_3(2 dx^+dx^- -\mu^2{\sum_{i=1}^{4}}x_i^2(dx^+)^2+
(dx_5)^2+(dx_6)^2)\cr
& \cr
&+& f^{1\over2}_3{\sum_{a=1..4,7,8}}(dx_a)^2  \cr
& \cr
H_{+12}&=&H_{+34}= 2\mu, \cr
& \cr
F_{+ - 5 6 a}&=& \partial_a {f_3}^{-1},~~~~e^{2\phi} = 1,
\label{d3-bkgrd}
\end{eqnarray}
with $f_3$ being the harmonic function in the transverse space of the
$D3$-brane.

Now we present the supergravity solution of intersecting 
($Dp-Dp^{\prime}$)-brane system in PP-wave background. These solutions are
described as `branes lying within branes'. In particular for $D1 - D5$ 
case, the solution is given by:
\begin{eqnarray}
ds^2&=&(f_1 f_5)^{-{1\over 2}}(2 dx^+dx^-
-\mu^2{\sum_{i=1}^{4}}x_i^2 (dx^+)^2) + \Big({f_1 \over f_5}\Big)^{1\over
  2}{\sum_{a=5}^{8}}(dx^a)^2 \cr
& \cr
&+&(f_1 f_5)^{1\over 2}{\sum_{i=1}^{4}}(dx_i)^2, \cr
& \cr
e^{2\phi}&=&f_1\over f_5, \cr
& \cr
H_{+ 1 2}&=&H_{+ 3 4} = 2\mu,\cr
& \cr
F_{+ - i}&=&\partial_{i}{f_1}^{-1},~~~~F_{m n p} = \epsilon_{m n p l}
\partial_{l}f_5,
\label{d1-d5}
\end{eqnarray}  
with $f_1$ and $f_5$ satisfying the Green function equations for $D1$ and 
$D5$-branes respectively. 
Eqn. (\ref{d1-d5}) provides one of the
main result of our paper and shows that, as in the flat space,
intersecting brane solutions are possible in the PP-wave background
as well. We have once again checked that the solution
presented above do satisfy type IIB field equations of motion.

One can now apply T-duality transformations to generate more intersecting 
brane solutions starting from the one given in eqn. (\ref{d1-d5})
\cite{myers,kamal}. Note that the directions $x^5,..,x^8$ are
transverse to the $D$-string in eqn. (\ref{d1-d5}), 
whereas they lie along the longitudinal directions of $D5$. 
As a result, one can easily obtain solutions of the type 
$D2-D4$ as well as $D3-D3^{\prime}$ in this PP-wave
background. These solutions will give a  
PP-wave generalization of the intersecting solutions given in 
\cite{tseyt6}. We however skip the details of this analysis.

\subsect{\bf $ p-p^{\prime}$ Branes in R-R PP-Wave Background}

In this section, we will present the $Dp$ as well as
$(Dp-Dp^{\prime})$-branes in R-R PP-wave of $AdS_3\times
S^3\times R^4$. 
These backgrounds can be  obtained from the solutions given in 
the last subsection by applying
$S$ and $T$-duality transformations in several steps.
For example, from the $D3$-brane solution in NS-NS PP-wave background
(\ref{d3-bkgrd}), one gets a $D3$-brane in RR PP-wave
background under $S$-duality transformation. 
Now applying $T$-duality along the directions ($x^5, x^6$), 
we can generate a $D$-string solution. 
On the other hand, by applying $T$-duality along two transverse 
directions, $(x^7, x^8)$, of the $D3$-brane, 
one gets a $D5$-brane lying along $(x^+, x^-, x^5,..,x^8)$ directions.
Supergravity solution of a system of  
$N$ $D$-strings is then given explicitly by:
\begin{eqnarray}
ds^2&=&f^{-{1\over2}}_1(2 dx^+dx^- - \mu^2{\sum_{i=1}^{4}}x_i^2(dx^+)^2)
+ f^{1\over2}_1{\sum_{a=1}^{8}}(dx^a)^2  \cr
& \cr
e^{2\phi}&=& f_1,~~~~F_{+1256} = F_{+3456}= 2\mu, \cr
& \cr
F_{+-a}&=&\partial_a f^{-1}_1,~~~~f_1 = 1 + {N g_s l^6_s\over r^6},
\label{d1rr-ads3}
\end{eqnarray}
with $f_1$ satisfying the Green function in $8$-dimensional transverse
space. One notices that the solution has a constant $5$- form field
strength.

The supergravity solution of $D5$-brane is given by:
\begin{eqnarray}
ds^2&=&f^{-{1\over2}}_5(2 dx^+dx^- - \mu^2{\sum_{i=1}^{4}}x_i^2(dx^+)^2
 + {\sum_{a=5}^{8}}(dx^a)^2) + 
f^{1\over2}_5{\sum_{i=1}^{4}}(dx^i)^2  \cr
& \cr
e^{2\phi}&=& f^{-1}_5,~~~~F_{+1256} = F_{+3456} = F_{+1278} =
F_{+3478} = 2\mu, \cr
& \cr
F_{mnp}&=&\epsilon_{m n p q}\partial_q f_5,
~~~~f_5 = 1 + {N g_s l^2_s\over r^2},
\label{d5rr-ads3}
\end{eqnarray}
with $f_5$ satisfying the Green function in the tranverse directions
$(x^1,...,x^4)$.
Now we will present the $(Dp-Dp^{\prime})$-brane solutions in RR
PP-wave background. In particular, to write down the supergravity
solution of a $(D1-D5)$ system, we made an ansatz which combines 
the $D$-string of eqn. (\ref{d1rr-ads3}) and $D5$-brane given 
in eqn. (\ref{d5rr-ads3}). The final configuration is as follows:
\begin{eqnarray}
ds^2&=&(f_1 f_5)^{-{1\over 2}}(2 dx^+dx^-
-\mu^2{\sum_{i=1}^{4}}x_i^2 (dx^+)^2) + \Big({f_1 \over f_5}\Big)^{1\over
  2}{\sum_{a=5}^{8}}(dx^a)^2 \cr
& \cr
&+&(f_1 f_5)^{1\over 2}{\sum_{i=1}^{4}}(dx_i)^2, \cr
& \cr
e^{2\phi}&=&f_1\over f_5, \cr
& \cr
F_{+ 1 2 5 6}&=&F_{+ 3 4 5 6} = F_{+ 1 2 7 8} = F_{+ 3 4 7 8} = 2\mu,\cr
& \cr
F_{+ - i}&=&\partial_{i}{f_1}^{-1},~~~~F_{m n p} = \epsilon_{m n p l}
\partial_{l}f_5,
\label{d1-d5-rr}
\end{eqnarray}  
with $f_1$ and $f_5$ being the Green function in the common tranversee
space. One can check that the solution presented above do satisfy the
type IIB field equations. Once again, more $p-p^{\prime}$ branes
can be obtained from the $D1-D5$ solution in (\ref{d1-d5-rr}) by
applying $T$-dualities. 

One may also attempt to find a $(D1-D5)$ solution by 
taking a decoupling limit, followed by the Penrose scaling, 
of the solution presented 
in\cite{papa} in a similar way as the $D5$-brane solution in\cite{kumar}.
The starting solution along which one would take the Penrose limit
is as follows:
\begin{eqnarray}
ds^2&=&{1\over{({H^{\prime}_1}H^{\prime}_5)^{1/2}}}[{r^2\over {R^2_1}}(-dt^2
+ dx^2)] + \Big({H^{\prime}_1\over H^{\prime}_5}\Big)^{1/2}{R^2_1\over 
  r^2} dr^2\cr 
& \cr
&+&\Big({H^{\prime}_1\over H^{\prime}_5}\Big)^{1/2}(d\psi^2 + \sin^2\psi
d\Omega^2_2) + (H^{\prime}_1 H^{\prime}_5)^{1/2} (dy^2 + y^2
d\Omega^2_3),
\end{eqnarray}
where 
\begin{equation}
H_1 = 1 + {R_1^2\over x^2},~~~H_5 = 1 + {R_5^2\over x^2},~~~
H^{\prime}_1 = 1 + {{R^{\prime}}_1^2\over y^2},
~~~H^{\prime}_5 = 1 + {{R^{\prime}}_5^2\over y^2}.
\end{equation}  
One however notices that different terms in the metric above
come with different powers of $H^{\prime}_1$,
leading to difficulty in choosing a `null geodesic' to define
an appropriate Penrose limit and find brane solutions.

\sect{\bf Supersymmetry Analysis}

\subsect{\bf NS-NS PP-wave}

In this section we will present the supersymmetry of the solutions
described earlier in section-(2.1). 
The supersymmetry variation of dilatino and 
gravitino fields of type IIB supergravity in ten dimension, 
in string frame, is given by\cite{schwarz,fawad,alishah}:
\begin{eqnarray}
\delta \lambda_{\pm} &=& {1\over2}(\Gamma^{\mu}\partial_{\mu}\phi \mp
{1\over 12} \Gamma^{\mu \nu \rho}H_{\mu \nu \rho})\epsilon_{\pm} + {1\over
  2}e^{\phi}(\pm \Gamma^{M}F^{(1)}_{M} + {1\over 12} \Gamma^{\mu \nu
  \rho}F^{(3)}_{\mu \nu \rho})\epsilon_{\mp},
\label{dilatino}
\end{eqnarray}
\begin{eqnarray}
\delta {\Psi^{\pm}_{\mu}} &=& \Big[\partial_{\mu} + {1\over 4}(w_{\mu
  \hat a \hat b} \mp {1\over 2} H_{\mu \hat{a}
  \hat{b}})\Gamma^{\hat{a}\hat{b}}\Big]\epsilon_{\pm} \cr
& \cr
&+& {1\over 8}e^{\phi}\Big[\mp \Gamma^{\mu}F^{(1)}_{\mu} - {1\over 3!}
\Gamma^{\mu \nu \rho}F^{(3)}_{\mu \nu \rho} \mp {1\over 2.5!}
\Gamma^{\mu \nu \rho \alpha \beta}F^{(5)}_{\mu \nu \rho \alpha
  \beta}\Big]\Gamma_{\mu}\epsilon_{\mp},
\label{gravitino}
\end{eqnarray}
where we have used $(\mu, \nu ,\rho)$ to describe the ten
dimensional space-time indices, and hat's represent the corresponding
tangent space indices.
Solving the above two equations for the solution describing a D-string
as given in eqn. (\ref{d1-ads3}), we get several conditions. First, the   
dilatino variation gives:
\begin{eqnarray}
\Gamma^{\hat{a}}\epsilon_{\pm} -
\Gamma^{\hat{+}\hat{-}\hat{a}}\epsilon_{\mp}=0,
\label{dilt-1}
\end{eqnarray}
\begin{eqnarray}
(\Gamma^{\hat{+}\hat{1}\hat{2}}+\Gamma^{\hat{+}\hat{3}\hat{4}})
\epsilon_{\mp}=0.
\label{dilt-2} 
\end{eqnarray}
In fact, both the conditions (\ref{dilt-1}) and (\ref{dilt-2}), are required
for satisfying dilatino variation condition.
Gravitino variation gives the following conditions on the spinors:
\begin{eqnarray}
\delta \psi_+^{\pm} &\equiv &\partial_{+}\epsilon_{\pm}\mp{\mu\over2}
f^{-{1\over2}}_1(\Gamma^{\hat{1}\hat{2}}+\Gamma^
{\hat{3}\hat{4}})\Gamma^{\hat{-}}\epsilon_{\pm}=0,\>\>\>
\delta \psi_-^{\pm} \equiv \partial_{-}\epsilon_{\pm}=0,\cr
& \cr 
\delta \psi_a^{\pm}&\equiv &\partial_{a}\epsilon_{\pm}
= - {1\over 8} {f_{1,a}\over f_1} \epsilon_{\pm},\>\>\>
\delta \psi_i^{\pm} \equiv \partial_{i}\epsilon_{\pm}=
- {1\over 8} {f_{1,i}\over f_1} \epsilon_{\pm}.
\label{1_i}
\end{eqnarray}
In writing the above set of equations, we have also 
imposed a necessary condition:
\begin{eqnarray}
\Gamma^{\hat +}\epsilon_{\pm} = 0,
\end{eqnarray} 
in addition to (\ref{dilt-1}). Further, by using
\begin{eqnarray}
(1 - \Gamma^{\hat 1\hat 2\hat 3\hat 4})\epsilon_{\pm} = 0,
\end{eqnarray}
all the supersymmetry conditions are solved by spinors:
$\epsilon_{\pm} = exp(-{1\over 8} ln f_1)\epsilon^0_{\pm}$, with
$\epsilon^0_{\pm}$ being a constant spinor.
$D$-string solution in eqn. (\ref{d1-ads3}) therefore preserves $1/8$ 
supersymmetry. All other $Dp$-branes ($p=1,..,5$), obtained by 
applying T-dualities as discussed above, will also preserve same 
amount of supersymmetry. 

Next, we will analyze the supersymmetry properties of the intersecting
branes. We will concentrate on the $(D1-D5)$-case explicitly.
The dilatino variation gives the following conditions on the spinors:
\begin{eqnarray}
\Gamma^{\hat i}\epsilon_{\pm} - \Gamma^{\hat +\hat -\hat
  i}\epsilon_{\mp} = 0,
\label{d1-d5-dila1}
\end{eqnarray}
\begin{eqnarray}
\Gamma^{\hat i}\epsilon_{\pm} + {1\over 3!}\epsilon_{\hat i\hat j\hat
  k\hat l}\Gamma^{\hat j\hat k\hat l}\epsilon_{\mp} = 0,
\label{d1-d5-dila2}
\end{eqnarray}
\begin{eqnarray}
(\Gamma^{\hat +\hat 1\hat 2} + \Gamma^{\hat + \hat 3 \hat
  4})\epsilon_{\mp} = 0.
\label{d1-d5-dila3}
\end{eqnarray}
One needs to impose all the three conditions, specified above for the
dilatino variation to vanish.
On the other hand, the gravitino variation gives:
\begin{eqnarray}
\delta\psi_+^{\pm}&\equiv &\partial_+ \epsilon_{\pm} \mp {\mu\over
  2}(f_1 f_5)^{-{1\over 2}}(\Gamma^{\hat 1\hat 2} + \Gamma^{\hat 3 \hat
  4})\Gamma^{\hat -}\epsilon_{\pm} = 0,\>\>\> 
\delta\psi_-^{\pm} \equiv \partial_- \epsilon_{\pm} = 0,\cr
& \cr
\delta\psi_i^{\pm}&\equiv &\partial_i \epsilon_{\pm} = 
 - {1\over 8} \left[{f_{1,a}\over f_1} \epsilon_{\pm} +
{f_{5,a}\over f_5}\right] \epsilon_{\pm} ,\>\>\>
\delta\psi_a^{\pm} \equiv \partial_a \epsilon_{\pm} = 0.
\end{eqnarray}
In writing down the above gravitino variations we have once again
made use of the projection $\Gamma^{\hat +}\epsilon_{\pm} = 0$.
The above set of equations can be solved by imposing: 
\begin{eqnarray}
(1 - \Gamma^{\hat 1\hat 2\hat 3\hat 4})\epsilon_{\pm} = 0,
\end{eqnarray} 
in addition to (\ref{d1-d5-dila1}) and the solution is given as:
$\epsilon_{\pm} = exp(-{1\over 8} ln (f_1f_5))\epsilon^0_{\pm}$.
One therefore has $1/8$
supersymmetry for the $(D1-D5)$ solution presented in eqn. (\ref{d1-d5}).

\subsect{\bf R-R PP-wave}
Now, we will present the supersymmetry of the $Dp$ as well as
$Dp-Dp^{\prime}$ branes in R-R PP-wave background given in 
section-(2.2). First we will discuss the
supersymmetry of the $D$-string in
eqn. (\ref{d1rr-ads3}). The dilatino variation (\ref{dilatino}),
gives:
\begin{eqnarray}
\Gamma^{\hat a}\epsilon_{\pm} - \Gamma^{\hat +\hat -\hat a} 
\epsilon_{\mp} = 0.
\label{d1rr-dila}
\end{eqnarray}  
Gravitino variation gives the following conditions on the spinors:
\begin{eqnarray}
\delta \psi_+^{\pm} &\equiv &\partial_{+}\epsilon_{\pm}\mp{\mu\over8}
f^{-{1\over2}}_1\left((\Gamma^{\hat +\hat 1\hat 2\hat{5}\hat{6}}+\Gamma^
{\hat +\hat 3\hat 4\hat{5}\hat{6}}) +(\Gamma^{\hat +\hat 1\hat
  2\hat{7}\hat{8}}+\Gamma^{\hat +\hat 3\hat 4\hat{7}\hat{8}})\right)
\Gamma^{\hat{-}}\epsilon_{\pm}=0,\cr
& \cr
\delta \psi_-^{\pm} &\equiv& \partial_{-}\epsilon_{\pm}=0,\cr
& \cr 
\delta \psi_a^{\pm}&\equiv& \partial_{a}\epsilon_{\pm}
= - {1\over 8} {f_{1,i}\over f_1} \epsilon_{\pm},\>\>\>
\delta \psi_i^{\pm} \equiv \partial_{i}\epsilon_{\pm}
= - {1\over 8} {f_{1,a}\over f_1} \epsilon_{\pm},
\label{d1rr-susy}
\end{eqnarray}
where we have once again imposed a
necessary condition: $\Gamma^{\hat +}\epsilon_{\pm} = 0$, in addition
to (\ref{d1rr-dila}). Further, by using the condition:
\begin{eqnarray}
(1 - \Gamma^{\hat 1\hat 2\hat 3\hat 4}) = 0,
\end{eqnarray}
all the supersymmetry conditions are satisfied, thus 
preserving $1/8$ unbroken supersymmetry.

Next, we present the supersymmetry property of the 
$D1-D5$ solution written in eqn. (\ref{d1-d5-rr}).
The dilatino variation (\ref{dilatino}) gives the following 
conditions on spinors:
\begin{eqnarray}
\Gamma^{\hat i}\epsilon_{\pm} - \Gamma^{\hat +\hat -\hat
  i}\epsilon_{\mp} = 0,
\label{d1-d5rr-dila1}
\end{eqnarray}
\begin{eqnarray}
\Gamma^{\hat i}\epsilon_{\pm} + {1\over 3!}\epsilon_{\hat i\hat j\hat
  k\hat l}\Gamma^{\hat j\hat k\hat l}\epsilon_{\mp} = 0.
\label{d1-d5rr-dila2}
\end{eqnarray}  
On the other hand, the gravitino variation (\ref{gravitino}), gives the
following conditions:
\begin{eqnarray}
\delta\psi_+^{\pm}&\equiv &\partial_+ \epsilon_{\pm} \mp {\mu\over
  8}(f_1 f_5)^{-{1\over 2}}\left((\Gamma^{\hat +\hat 1\hat 2\hat 5\hat 6}
+ \Gamma^{\hat +\hat3 \hat 4 \hat 5\hat 6}) 
+(\Gamma^{\hat +\hat 1\hat 2\hat 7\hat 8}
+\Gamma^{\hat +\hat3 \hat 4 \hat 7\hat 8})\right) 
\Gamma^{\hat -}\epsilon_{\pm} = 0,\cr
& \cr 
\delta\psi_-^{\pm} &\equiv& \partial_- \epsilon_{\pm} = 0,\>\>\>
\delta\psi_a^{\pm} \equiv \partial_a \epsilon_{\pm} = 0,\cr
& \cr
\delta\psi_i^{\pm}&\equiv& \partial_i \epsilon_{\pm} = 
- {1\over 8} \left[{f_{1,i}\over f_1} + {f_{5,i}\over f_5}\right] 
\epsilon_{\pm},
\end{eqnarray}
where we have once again used a necessary condition: $\Gamma^{\hat
  +}\epsilon_{\pm} = 0$ along with the ones in 
(\ref{d1-d5rr-dila1}) and (\ref{d1-d5rr-dila2}).
The above set of equations can be solved 
by imposing further:
\begin{eqnarray}
(1 - \Gamma^{\hat 1\hat 2\hat 3\hat 4})\epsilon_{\pm} = 0. 
\end{eqnarray}
One therefore has $1/8$ supersymmetry for the $D1-D5$ system described
in eqn. (\ref{d1-d5-rr}) as well.

\sect{\bf Worldsheet Construction of $p-p^{\prime}$ Branes}

\subsect{\bf NS-NS PP-wave}

In this section, we will discuss the $(D1-D5)$-brane system, 
constructed earlier in the paper, from 
the point of view of first quantized string theory in Green-Schwarz
formalism, in light-cone gauge. In the present case, in flat
directions $x_{\alpha} (\alpha = 5,...,8)$, we have the Dirichlet boundary
condition at one end and Neumann boundary condition at 
the other end of the open string.  
Along $x_i$ $(i = 1,...,4)$ directions, one has the usual 
Dirichlet boundary condition. The relevant classical action to be
studied in our case (after imposing the light-cone gauge conditions
on fermions and bosons \cite{tseytlin} )
is as follows \cite{michi}:
\begin{eqnarray}
L=L_b + L_f,
\end{eqnarray}
where
\begin{eqnarray}
L_b&=&\partial_+ u\partial_- v - m^2 x^2_i + \partial_+ x_i\partial_-
x_i + \partial_+ x_{\alpha}\partial_- x_{\alpha}\cr
& \cr
&+& \mu \sum_{(i,j) = (1, 2), (3, 4)} x^i (\partial_+ u \partial_- x^j
- \partial_- u \partial_+ x^j ), 
\end{eqnarray}
\begin{eqnarray}
L_f = i S_R (\partial_+ - m M )S_R 
+ i S_L (\partial_-  + m M )S_L,
\end{eqnarray}
with
\begin{eqnarray}
m\equiv \alpha^{\prime}p^{u}\mu = 2\alpha^{\prime}p_v \mu, 
\end{eqnarray}
\begin{eqnarray}
M = -{1\over 2}(\gamma^{1 2} + \gamma^{3 4}).
\end{eqnarray}
Eight component real spinors $(S_L, S_R)$ have been obtained from 
16-component Majorana-Weyl spinors in the left and the right sector
after solving the light-cone gauge conditions.

The equations of motion and boundary conditions for bosons 
$x^i, x^{\alpha}$ in our case are as follows:
\begin{eqnarray}
\partial_+\partial_- x_{i_1} + m^2x_{i_1} - 
m \epsilon^{i_1 j_1} (\partial_- x^{j_1} - \partial_+ x^{j_1})
= 0,~~~~~~~
\partial_+\partial_-x_{\alpha} = 0,
\end{eqnarray}  
\begin{eqnarray}
\partial_{\sigma} x^{\alpha}{\Big|_{\sigma =0}} = \partial_{\tau}
x^{\alpha}{\Big|_{\sigma =\pi}} = 0,\;\;\;
x^i {\Big|_{\sigma =0,\pi}}=  constant. 
\end{eqnarray}
The solutions to the bosonic equations of motion, with the boundary
conditions specified above, 
is given by (defining $X^{\hat{1}} = {1\over \sqrt 2} (x^1 + i x^2)$ and  
$X^{\hat{2}} = {1\over \sqrt 2} (x^3 + i x^4)$) \cite{michi} : 
\begin{eqnarray}
X^{\alpha}(\sigma,\tau) = i \sum_{r\in (z + {1\over 2})}
{1\over r}\alpha_r^{\alpha} e^{-i r \tau}\cos r\sigma,\;\; 
(\alpha = 5,..,8),
\label{long-boson}
\end{eqnarray} 
\begin{eqnarray}
X^{\hat{i}}(\sigma,\tau) = e^{-2 i m\sigma}\left[x_0 +(x_1 e^{2 i
  m\sigma}-x_0){\sigma\over \pi} + i \sum_{n\ne 0}{1\over n}\alpha_n^i
e^{-i n \tau}\sin n\sigma\right].
\label{trans-boson}
\end{eqnarray}

To consider the equations of motion of fermions, we note that 
the matrix $M$ evidently breaks the $SO(8)$ symmetry further, 
and thereby splits the fermions in the $8\rightarrow 4 + 4$ way:
$S_L\rightarrow(\tilde S_L,\hat S_L), S_R\rightarrow(\tilde S_R,\hat
S_R)$:
\begin{eqnarray}
\gamma^{1234} \pmatrix{\tilde{S}_{L,R} \cr \hat{S}_{L,R}} = 
\pmatrix{- \tilde{S}_{L,R} \cr \hat{S}_{L,R}}. \label{ga1234}
\end{eqnarray}
In this connection, one also introduces $4\times 4$ matrices 
$\Lambda$ and $\Sigma$:
\begin{eqnarray}
\gamma^{12}\pmatrix{\tilde{S}_{L,R} \cr \hat{S}_{L,R}} = 
-\pmatrix{\Lambda \tilde{S}_{L,R} \cr\Sigma \hat{S}_{L,R}}, \label{gamma12}
\end{eqnarray}
with $\Lambda^2 = \Sigma ^2 = -1$. 
$\Lambda$ and $\Sigma$ in the above equation are $4\times 4$ 
antisymmetric matrices with eigenvalues $\pm i$. Using these 
notations, one has:
\begin{eqnarray}
M \pmatrix{\tilde{S}_{L,R} \cr \hat{S}_{L,R}} 
= \pmatrix{\Lambda \tilde{S}_{L,R} \cr  0}.
\end{eqnarray}
The equations of motion written in terms of $(\tilde{S}_L,{\hat{S}}_L)$ and 
$(\tilde{S}_R,{\hat{S}}_R)$ are then of the form: 
\begin{equation}
\partial_{+}(e^{2 m\tau} \tilde{S}_{R}) =0, \>\>\>\>
\partial_{-}(e^{- 2m\tau }\tilde{S}_{L}) =0,
\label{long-ferm}
\end{equation}
\begin{equation}
\partial_{+}\hat{S}_{R} =0, \>\>\>\>
\partial_{-}\hat{S}_{L} =0.
\label{trans-ferm}
\end{equation}
Now we will write down the boundary conditions for the fermions
in the mixed sector. As the equations of motion and the 
boundary condition for the components $\hat{S}_{L,R}$ are identical to the
ones in flat space, we only concentrate on finding explicit solution for 
$\tilde{S}_{L,R}$ below.
Following\cite{tseytlin,kumar,michi}, one can write down
the boundary conditions for the fermions as:
\begin{eqnarray}
\tilde{S_L}\big|_{\sigma=0} = - \tilde{S_R}\big|_{\sigma=0},\>\>\>\>
\end{eqnarray}
\begin{eqnarray}
\tilde{S_L}\big|_{\sigma=\pi} = \tilde{S_R}\big|_{\sigma=\pi},\>\>\>\>
\end{eqnarray}
\begin{eqnarray}
\hat{S_L}\big|_{\sigma=0,\pi} = \hat{S_R}\big|_{\sigma=0,\pi}.
\end{eqnarray}
The solution for $\tilde{S}_{L,R}$ equations of motion
(\ref{long-ferm}), with the above boundary condition,
can be read from \cite{michi}, and has the following form:
\begin{eqnarray}
\tilde{S}_L = - e^{-2 m\sigma\Lambda}\sum_{r\in (z + {1\over 2})}s_r
e^{-i r(\tau + \sigma)}, 
\end{eqnarray}
\begin{eqnarray}
\tilde{S}_R =  e^{-2 m\sigma\Lambda}\sum_{r\in (z + {1\over 2})}s_r
e^{-i r(\tau - \sigma)}. 
\end{eqnarray}

The canonical quantization conditions as well as the worldsheet 
hamiltonian for the $D1-D5$ system discussed above can also be
written in a straightforward manner following the procedure in 
\cite{michi}. We skip these details. 

\subsect{\bf R-R PP-wave}
Now, we present the worldsheet analysis of the $(D1-D5)$-system
discussed earlier in section-(2.2). This can be done by 
realizing that the PP-wave background for these solutions 
is given by a $T$-dual configuration of the ones presented
in \cite{tseytlin,kumar}. More explicitly, in the worldsheet
action in the present case:
\begin{eqnarray}
L=L_B + L_F,
\end{eqnarray}
where
\begin{eqnarray}
L_B = \partial_+ u\partial_- v - m^2 x^2_i + \partial_+ x_i\partial_-
x_i + \partial_+ x_{\alpha}\partial_- x_{\alpha}, 
\end{eqnarray}
\begin{eqnarray}
L_F = i {S}_R \partial_+ {S}_R 
+ iS_L \partial_+ S_L - 2im {S}_L 
M {S}_R,
\end{eqnarray}
with
\begin{eqnarray}
m\equiv \alpha^{\prime}p^{u}\mu = 2\alpha^{\prime}p_v \mu, 
\end{eqnarray}
\begin{eqnarray}
M = -{1\over 2}(\gamma^{1 2} + \gamma^{3 4}) \gamma^{5 6},\>\>\>
\label{Matrix}
\end{eqnarray}
the terms involving fermions are easily seen to be related to the
ones in \cite{tseytlin,kumar} through $T$-dualities along 
$x^5$ and $x^6$. This in fact leads to the relation: 
\begin{eqnarray}
{S^{\prime}}_R = \gamma^{5 6}S_R,
\label{t-fermion}
\end{eqnarray}
and reproduces the original action in \cite{tseytlin}. The mode 
expansion for fermions as well as canonical quantization conditions
can therefore be also written down in a straightforward manner. 
We end this section by pointing out that, since the $D$-brane solutions
found in this paper are preserving less than $1/2$ supersymmetry,
some of the restrictions on the brane directions, imposed using 
zero mode considerations\cite{dabh} do not directly apply above. 

\sect{\bf Branes in `Little String Theory'}

\subsect{\bf Supergravity Backgrounds} 
In this section, we discuss the branes in the Penrose limit of 
`little string theory'(LST). PP-waves of non-local theories have been
discussed recently in the literature\cite{rangamani,sakai,kumar1}.
Among them, `little string theory' arises on the world volume of $NS5$-brane
when a decoupling limit, $g_s\rightarrow 0$ with fixed
$\alpha^{\prime}$, is taken\cite{berkooz,seiberg}. To construct our
solution, we start with the NS5-brane solution given by the metric and
dilaton:
\begin{eqnarray}
ds^2 &=& -dt^2 + dy^{2}_5 + H(r)(dr^2 + r^2d\Omega^{2}_3),\cr
& \cr
e^{2\Phi} &=& g^{2}_s H(r),
\end{eqnarray}  
with $H(r) = 1 + {Nl^{2}_s \over r^2}$. 
The near horizon limit of the above solution is 
the linear dilaton geometry, which in the string frame is given by,
\begin{eqnarray}
ds^2= Nl^{2}_s\big(-d\tilde{t}^2 + \cos^2\theta d\psi^{2} + d\theta^{2}
+ \sin^2\theta d\phi^{2} +{dr^2 \over r^2}\big) + dy^{2}_5,
\label{little-1}
\end{eqnarray}
with $t = \sqrt{N}l_s \tilde{t}$. PP-wave background of LST is then
found by applying Penrose limit to (\ref{little-1}). We however
consider the case after applying S-duality on eqn. (\ref{little-1}). 
Applying S-duality transformation and then taking Penrose limit
(as described in\cite{rangamani}), the background solution is given by: 
\begin{eqnarray}
ds^2&=& -4dx^{+}dx^{-} - \mu^{2}\vec{z}^{2}{dx^+}^2 + (d\vec{z})^2 +
dx^2 + dy^{2}_5 \cr
& \cr
e^{2\phi}&=&Const, \cr
& \cr
F_{+ 1 2}&=&C \mu,
\end{eqnarray}  
where $F_{+ 1 2}$ is the 3-form field strength in the $\vec{z}$-plane.
Now we proceed to analyze the existence and stability
of branes in this background. The supergravity solution for a system of 
D5-branes in this background is given by:
\begin{eqnarray}
ds^2&=&f^{-{1\over2}}\Big(-4 dx^+dx^- -\mu^2\vec{z}^2(dx^+)^2+
{\sum_{i =1}^{2}}(dz_i)^2+{\sum_{p = 1}^{2}}(dy_p)^2\Big)\cr
& \cr
&+& f^{1\over2}{\sum_{a=1}^{4}}(dx_a)^2,  \cr
& \cr
e^{2\Phi}&=&f^{-1},~~~~~F_{+ 1 2} = 2\mu, \cr
& \cr
F_{m n p}&=&\epsilon_{m n p r}\partial_{r}f,~~~ f = 1 + {Ng_s l^2_s\over 
  r^2}.
\label{litt-d5}
\end{eqnarray}
One notices that the background has only one constant 3-form field
strength $(F_{+ 1 2})$. We have once again verified that the solution 
presented above satisfies type IIB field equations.

One can then write down (by applying S-duality on (\ref{litt-d5}))
the NS5-brane in a PP-wave background of the
`little string theory'\cite{rangamani} as:
\begin{eqnarray}
ds^2&=&-4 dx^+dx^- -\mu^2\vec{z}^2(dx^+)^2+
{\sum_{i =1}^{2}}(dz_i)^2+{\sum_{p = 1}^{2}}(dy_p)^2 
+f{\sum_{a=1}^{4}}(dx_a)^2 \cr
& \cr
e^{2\Phi}&=&f,~~~~~H_{+ 1 2} = 2\mu, \cr
& \cr
H_{m n p}&=&\epsilon_{m n p r}\partial_{r}f,
\end{eqnarray}
with $H$'s being the NS- sector 3-form field strengths. 

We will now consider the dilatino and gravitino
variation of the solution presented in eqn. (\ref{litt-d5})
to study the supersymmetry properties.
The dilatino variation gives equations:
\begin{eqnarray}
\Gamma^{\hat a}\epsilon_{\pm} + {1\over 3!}\epsilon_{\hat a \hat b
  \hat c \hat d}\Gamma^{\hat b \hat c \hat d}\epsilon_{\mp} = 0,
\label{lstdila1}
\end{eqnarray}
\begin{eqnarray}
\Gamma^{\hat + \hat1 \hat2}\epsilon_{\mp} =0.
\label{lstdila2}
\end{eqnarray}
The gravitino variation leads to the equations:
\begin{eqnarray}
\delta\Psi^{\pm}_+ \equiv \partial_{+}\epsilon_{\pm} + 
{\mu^2 z_{\hat i} \over 2}\Gamma^{\hat + \hat i}\epsilon_{\pm} 
+ {1\over 16}\mu^2 \vec{z}^2 {f,\hat a\over f^{3\over 2}}
\Gamma^{\hat +\hat a}\epsilon_{\pm} - 
{\mu \over 4}\Gamma^{\hat +\hat1\hat2}\Gamma^{-}\epsilon_{\mp}=0,
\end{eqnarray}
\begin{eqnarray}
\delta\Psi^{\pm}_{-}\equiv \partial_{-}\epsilon_{\pm}=0,
\end{eqnarray} 
\begin{eqnarray}
\delta\Psi^{\pm}_{i}\equiv\partial_{i}\epsilon_{\pm} - {\mu\over
  4}\Gamma^{\hat{+}\hat1\hat2}\delta_{i\hat{i}}\Gamma^{\hat{i}}
\epsilon_{\mp}=0.
\end{eqnarray}
\begin{eqnarray}
\delta\Psi^{\pm}_{p}\equiv\partial_{p}\epsilon_{\pm} - 
{\mu\over 4}\Gamma^{\hat + \hat 1\hat 2}
\delta_{p\hat{p}}\Gamma^{\hat{p}}\epsilon_{\mp}=0.
\end{eqnarray}
\begin{eqnarray}
\delta\Psi^{\pm}_{a}\equiv\partial_{a}\epsilon_{\pm}  + 
{1\over 8} {f,a\over f}\epsilon_{\pm} - {\mu\over
4}f^{1\over 2}\Gamma^{\hat{+}\hat1\hat2}\delta_{a\hat{a}}\Gamma^{\hat{a}}
\epsilon_{\mp}=0,
\end{eqnarray}
In writing these set of equations we have used the
condition (\ref{lstdila1}). Imposing the condition $\Gamma^{\hat
  +}\epsilon =0$, we further reduce them to:
\begin{eqnarray}
\partial_{+}\epsilon_{\pm} - {\mu \over 4}\Gamma^{\hat
  +\hat1\hat2}\Gamma^{\hat-}\epsilon_{\mp} = 0,
\label{d5-soln}
\end{eqnarray}
\begin{eqnarray}
\partial_{-}\epsilon_{\pm}=0,~~~~\partial_{i}\epsilon_{\pm} = 0,~~~~~
\partial_{p}\epsilon_{\pm} = 0,~~~\partial_{a}\epsilon_{\pm} 
= - {1\over 8} {f,a\over f}\epsilon_{\pm}. \label{last-eqn.}
\end{eqnarray}
Since eqns. (\ref{d5-soln}) and (\ref{last-eqn.})
are integrable ones, hence in this case we get 
$1/4$ supersymmetry. It will also be nice to give a worldsheet
construction for such $D$-branes.

\sect{\bf Conclusion}

In this paper we have presented several supersymmetric 
$Dp$ and $Dp-Dp^{\prime}$-brane configurations in 
PP-wave background and analyzed their supersymmetry properties. 
We have also presented $(D1-D5)$-brane construction from the
point of view of massive Green-Schwarz formalism in the light cone 
gauge in NS-NS and R-R PP-wave of $AdS_3\times S^3\times R^4$. It will
be interesting to study the gauge theory duals of the branes
presented in this paper by using operators such as `defects'. 
One could possibly also look at the black hole physics using the 
$(D1-D5)$ system presented here in an attempt to understand their properties.

\vskip .2cm
\noindent 
{\Large \bf Acknowledgement}
\vskip .2cm
\noindent
We thank Rashmi R. Nayak, Koushik Ray and Sanjay for useful discussions.

\end{document}